\documentclass[12pt,preprint]{aastex}

\shorttitle{Duration distributions}

\shortauthors{Qin et al.}

\begin{document}

\title{Duration distributions for different softness groups of gamma-ray bursts}

\author{Y.-P. Qin\altaffilmark{1,2}, A. C. Gupta\altaffilmark{3},
J. H. Fan\altaffilmark{1}, C.-Y. Su\altaffilmark{4}, R.-J.
Lu\altaffilmark{2}}

\altaffiltext{1}{Center for Astrophysics, Guangzhou University,
Guangzhou 510006, P. R. China; ypqin@gzhu.edu.cn}

\altaffiltext{2}{Physics Department, Guangxi University, Nanning
530004, P. R. China}

\altaffiltext{3}{Aryabhatta Research Institute of Observational
Sciences (ARIES), Manora Peak, Nainital - 263129, India}

\altaffiltext{4}{Department of Physics, Guangdong Industry
University, Guangzhou 510006, P. R. China}

\begin{abstract}

Gamma-ray bursts (GRBs) are divided into two classes according to
their durations. We investigate if the softness of bursts plays a
role in the conventional classification of the objects. We employ
the BATSE (Burst and Transient Source Experiment) catalog and
analyze the duration distributions of different groups of GRBs
associated with distinct softness. Our analysis reveals that the
conventional classification of GRBs with the duration of bursts is
influenced by the softness of the objects. There exits a
bimodality in the duration distribution of GRBs for each
group of bursts and the time position of the dip in the
bimodality histogram shifts with the softness parameter.
Our findings suggest that the conventional classification scheme
should be modified by separating the two well-known populations in
different softness groups, which would be more reasonable than doing
so with a single sample. According to the relation between the dip
position and the softness parameter, we get an empirical function
that can roughly set apart the short-hard and long-soft bursts: $SP
= (0.100 \pm 0.028) T_{90}^{-(0.85 \pm 0.18)}$, where $SP$ is the
softness parameter adopted in this paper.

\end{abstract}



\section{Introduction}

The bimodality in the duration distribution of GRBs
suggests that, the GRB events exist in short and long
duration classes, which are separated at 2 seconds [1]. In general,
the short duration bursts are harder and long duration bursts are
softer. In the hardness ratio vs. duration plot, the two classes are
seen to distribute in distinct domains [1, 2], where the hardness
ratio $HR_{32}$ concerned is defined as the ratio of the total
counts in the $100-300$ keV and $50-100$ keV energy range (later,
the hardness ratio is defined as the ratio of the fluence in the
$100-300$ keV range to the fluence in the $50-100$ keV range [3]).
When sources of both classes are combined, the hardness ratio is
obviously correlated to the duration, while for each of the two
classes alone the two quantities are not correlated at all [4]. This
in turn strongly suggests the existence of the two classes of GRBs.
The study of GRB classification is very important since different
classes might have different progenitors. It was proposed that the
long duration bursts are caused by the massive star collapsars
[5$-$7] and shot duration bursts are produced in the event of binary
neutron star or neutron star-black hole mergers [8, 9].

After the successful launch of the Swift satellite [10], a large
number of evidences favor the two progenitor proposal for GRBs. Long
duration bursts were found to be originated from star-forming
regions in galaxies [11], and in some of these, supernovae were
detected to accompany the bursts [12, 13]. On the other hand, short
duration bursts were detected in regions with lower star-formation
rates, with no evidence of supernovae to accompany them
[14$-$16].

Although GRBs are known to belong to the two distinct classes, the
membership of the classes is hard to establish due to the overlap of
the duration distributions, which makes the classification of the
bursts an unsettled issue. Authors of Ref. [17] showed that, the
bimodal distribution of GRBs can be well accounted for by two
overlapped lognormal distributions, which suggests that each of the
two GRB populations is likely to form a single peak modality and
there would be a sufficient number of bursts that are mis-classified
by simply applying the criterion $T_{90}=2s$. Authors of Ref. [18]
used a multivariate analysis to discriminate between distinct
classes of GRBs, and from the third BATSE catalog they found that
instead of two, there are three classes of GRBs, which triggered a
debate lasting until the present time.

In distinguishing the X-ray flashes (XRFs), X-ray rich bursts (XRRs)
and conventional gamma-ray bursts (CGRBs), authors of Ref. [19] used
the ratio between the X-ray fluence $S_X$ and the gamma-ray fluence
$S_\gamma$ as a softness parameter to divide them. Their results
seem to be quite satisfactory. We wonder, if the softness parameter
can play a role in separating the short-hard and long-soft
populations of GRBs from the BATSE catalog. This is the motivation
for our analysis given below.

The BATSE catalog has the largest sample of GRBs available until today.
We employ this sample in the following analysis since the total
number of GRBs in the sample is large enough for a meaningful
statistical analysis.

In Section 2, we investigate if the duration distribution depends on
the softness parameter. Based on this investigation, we shortly
explore a possible classification of GRBs with an empirical curve in
Section 3. The case of Swift is considered in Section 4. Discussion
and conclusions are presented in Section 5.

\section{Duration distributions of GRBs with different softness parameters}

Motivated by Ref. [19]'s work, we wonder if the softness of BATSE
bursts can play a role in separating the short-hard and long-soft
populations of GRBs. However, most of the BATSE bursts do not have
X-ray fluxes, so, we cannot define the softness of the bursts with
$S_X/S_\gamma$. Therefore, we turn to consider the ratio between the
fluence in the lowest energy band and that in higher energy bands,
as adopted in Ref. [20] to distinguish them. This parameter is no
longer applicable to distinguish XRFs, XRRs and CGRBs. Here, we
define the softness parameter of BATSE bursts by
\begin{equation}
SP\equiv\frac{f_1}{f_2+f_3+f_4},
\end{equation}
where $f_1$, $f_2$, $f_3$, and $f_4$ are the fluences of the first
($20-50$ keV), second ($50-100$ keV), third
($100-300$ keV), and fourth ($>300$ keV)
BATSE channels, respectively.

\begin{figure}[tbp]
\begin{center}
\includegraphics[width=5in,angle=0]{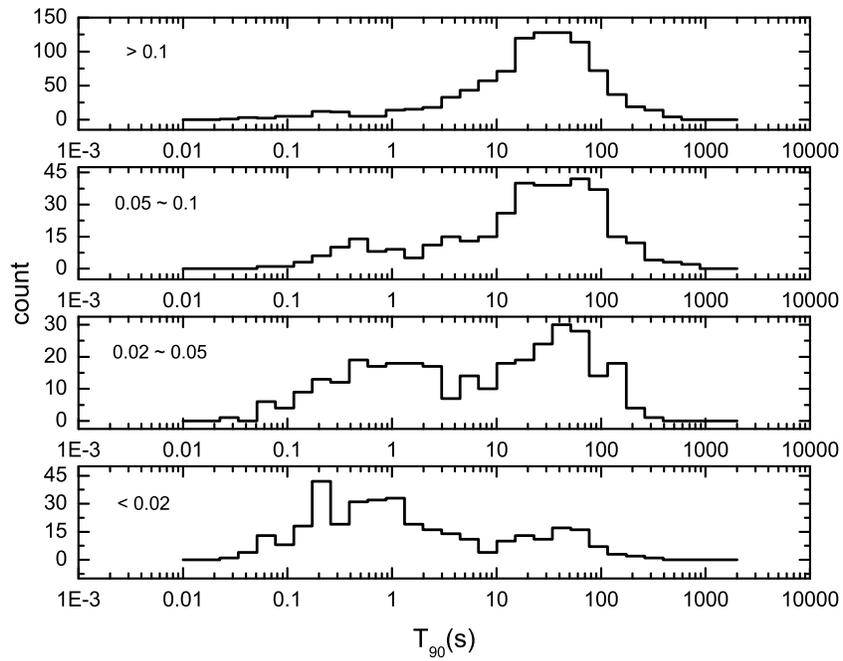}
\end{center}
\caption{Duration distributions of the BATSE bursts with different
softness parameters. Range of the adopted softness parameter is
mentioned at the upper left corner for each panel.} \label{Fig. 1}
\end{figure}

As hinted by Ref. [19, 20], we roughly divide the bursts into
several groups according to the softness parameter $SP$. To be
comparable to the conventional duration distribution plot, we divide
the duration range from 0.01 s to 2000 s in the logarithm format
into 30 bins. We require that the maximum count of each group should
be no less than 30 to keep a statistical significance. For the
softest group, we require that if there exist some short duration
bursts in this group, the corresponding count should be noticeable.
In this way, we get four groups of bursts. The distributions of
durations ($T_{90}$) of these groups are displayed in Figure 1, where
$T_{90}$ of a burst is the time interval during which 90\% of the
total observed counts have been detected. We find that the duration
histogram of each group bears a bimodality. The
positions of the dip lying between the two peaks of the
bimodality of different groups differ significantly,
where the softer the group, the smaller value of the duration
position of the dip. If separating the bursts into short and long
duration classes, then in different ranges of the softness parameter
the classes occupy significantly different percentage of the total
counts. The figure shows that the long duration bursts occupy a very
large percentage of the softest population ($SP \ge 0.1$), and they
cover a large range of durations, starting from $\sim 1$s to $\sim
1000$s. In contrast, short duration bursts occupy a very small
percentage of the softest population, and they cover a very small
range of durations. The situation changes for harder groups: the
percentage of long duration bursts becomes smaller and they cover
smaller ranges of durations, while the percentage of short duration
bursts becomes larger and they cover larger range of durations.

\begin{deluxetable}{lllll}
\tablewidth{0pt} \tablecaption{Parameters of different softness
groups.} \tablehead{  & \colhead{$\overline{SP}$} &
\colhead{$T_{90}(s)$} & \colhead{short$^a$} & \colhead{long$^b$} }
\startdata
 $  0.1  \leq SP    $    & $0.28 \pm 0.24$ &  $0.52^{+0.65}_{-0.42}$    &
          41&         895\\
 $  0.05 \leq SP < 0.1$  & $0.073 \pm 0.014 $ &   $0.82^{+1.52}_{-0.44}$    &
          42&         328\\
$  0.02  \leq SP < 0.05$ & $0.0338 \pm 0.0086 $ &
$3.3^{+5.2}_{-2.0}$    &
         135&         186\\
$             SP < 0.02$ & $0.0103 \pm 0.0056 $ &
$16.2^{+81.9}_{-3.2}$    &
         278&          67\\
\enddata
\tablenotetext{a}{The number of sort duration bursts with their
duration being no less than the $T_{90}$ value presented in the
third column.} \tablenotetext{b}{The number of long duration bursts
with their duration being larger than the $T_{90}$ value presented
in the third column.}
\end{deluxetable}

In the conventional classification of GRBs, a dip around 2 s in the
duration histogram plot in Figure 1a of Ref. [1] was proposed. For a
more precise measurement, the authors fitted a quadratic function
between the two peaks in the histogram and determined its minimum to
be $T_{90} = (1.2 \pm 0.4) s$. As an empirical analysis, we adopt
this quadratic function fitting method to locate the dip position
from the plots in Figure 1. In Table 1, dip positions and the numbers
of long and short duration bursts divided by this feature for
various softness groups are listed. In the first column, the softness
parameter range of each group is presented. The mean of the softness
parameter calculated with all the individual $SP$ values of the
bursts of the corresponding group is presented in the second column.
Positions of the dip are given in the third column. The numbers of
short and long duration bursts separated by the dip are presented in
the fourth and fifth columns respectively.

\section{Empirical curve roughly setting apart two populations of BATSE bursts}

A relation between the softness parameter and the dip position of
the bi-modal duration distribution of BATSE bursts hinted by Figure 1
can be evaluated from the data of Table 1. In Figure 2, we display the
result of a power-law analysis on the data. Fitting the data with a
power-law function yields (by performing a Spearman correlation
analysis with the ORIGIN software):
\begin{equation}
SP = (0.100 \pm 0.028) T_{90}^{-(0.85 \pm 0.18)}.
\end{equation}
The following are the statistical results of a linear correlation
analysis between $log SP$ and $log T_{90}$: the correlation
coefficient, -0.958; the number of data points, 4; the probability
of rejecting the null hypotheses, 0.0417. Owing to the limit of the
total counts, we have only 4 data points in Figure 2. The statistical
analysis performed here is thus not robust (we hence regard this
fitting curve as an empirical function
--- see the statement below). Note that one can divide the BATSE
catalog into more groups to have more data points. As a result, one
might find difficulties to locate the dip suggested in Figure 1 due to
data fluctuations.

\begin{figure}[tbp]
\begin{center}
\includegraphics[width=5in,angle=0]{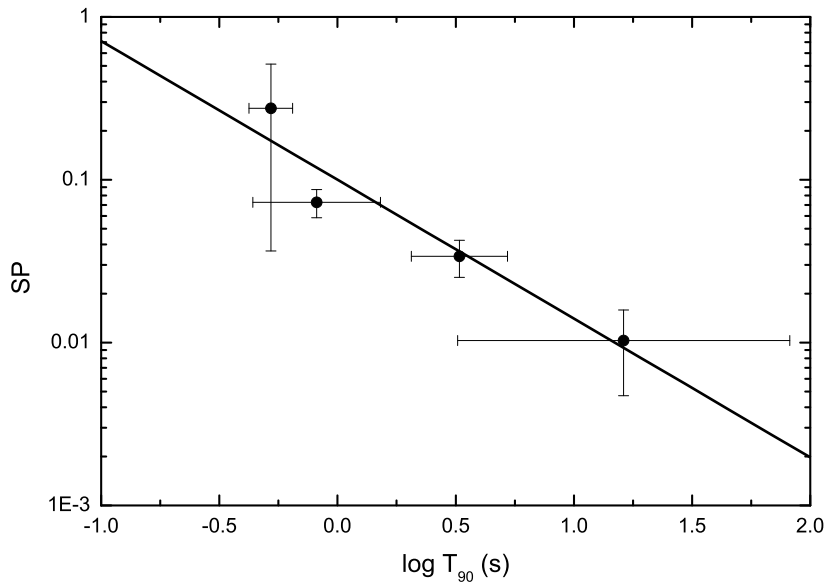}
\end{center}
\caption{Relation between the softness parameter and the position of
the duration distribution dip drawn from the BATSE catalog.}
\label{Fig. 2}
\end{figure}

\begin{figure}[tbp]
\begin{center}
\includegraphics[width=5in,angle=0]{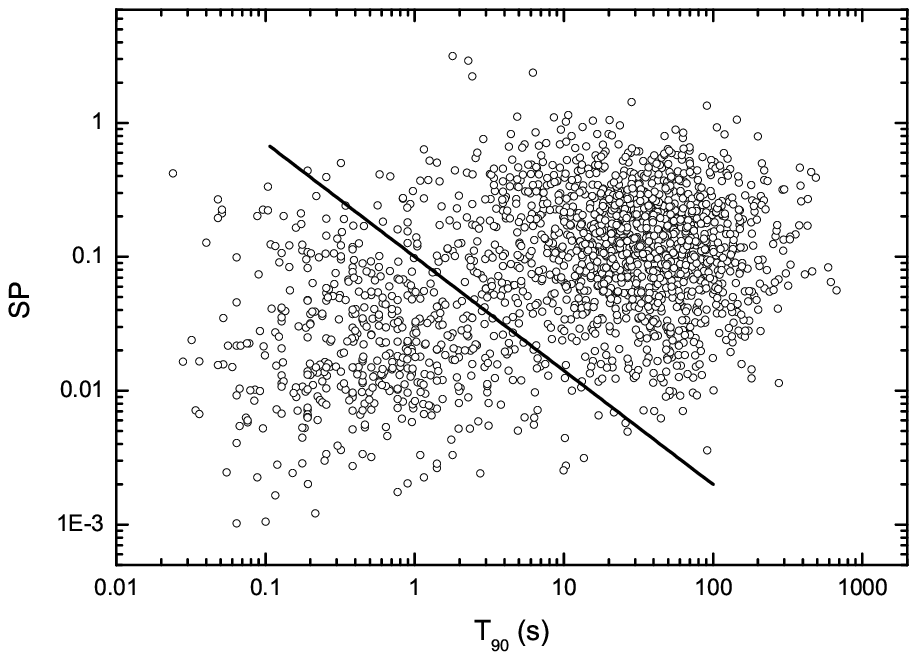}
\end{center}
\caption{Plot of the softness parameter vs. the duration of BATSE
bursts, where the solid line is the function of Equation (2).}
\label{Fig. 3}
\end{figure}

In Figure 3, we display the plot of the softness parameter vs. the
duration of BATSE bursts, where the function of Equation (2) is also
plotted. The figure shows that there exist two populations of bursts
which are clustered in two distinct domains in the plot. Between the
two clusters there is a ``gap'' with relatively sparse counts. The
function of Equation (2), which is the fitting curve in Figure 2, is
seen to pass through the ``gap'', roughly separating the two
populations. As the figure shows, most sources of the population
below the fitting curve are the conventional short bursts (the
duration is smaller than 2 s), and most sources of the population
above the fitting curve are the conventional long bursts (the
duration is larger than 2 s).

\begin{figure}[tbp]
\begin{center}
\includegraphics[width=5in,angle=0]{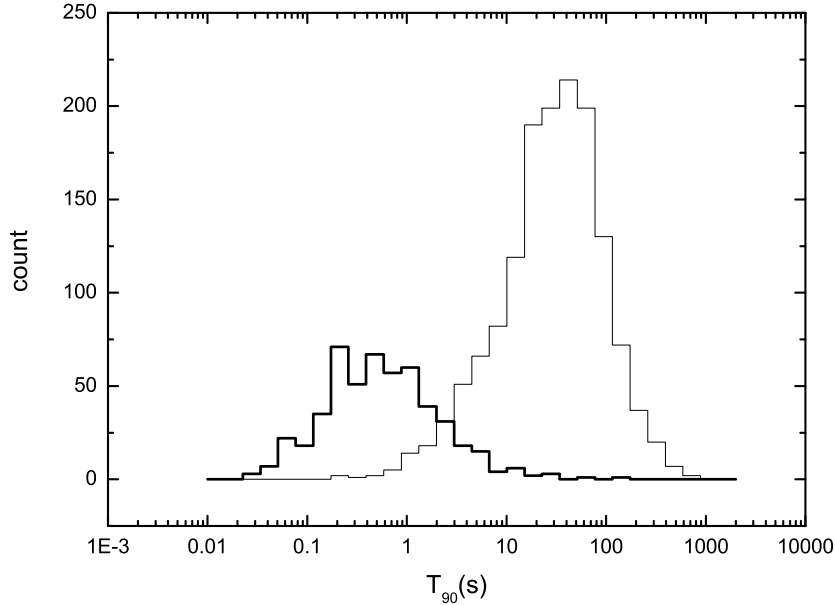}
\end{center}
\caption{Duration distributions of the two populations divided by
the curve of Equation (2), where the thick line represents the
short-hard population while the thin line denotes the long-soft
population drawn from the BATSE catalogue.} \label{Fig. 1}
\end{figure}

We find that: a) the number of the population below the fitting
curve is 511, and among them there are 433 bursts (84.7\% of the
population) with their duration satisfying $T_{90}\leq 2 s$; b) the
number of the population above the fitting curve is 1461, and among
them there are 1418 bursts (97.1\% of the population) with their
duration satisfying $T_{90} > 2 s$. The duration distributions of
the two populations divided by the fitting curve is shown in Figure 4.
Remind that the fitting curve represents the relation between the
dip positions of the duration distribution of bursts and the
softness parameter. According to this analysis (refer to Figures 1, 3
and 4), we regard the fitting curve as an empirical curve roughly
setting apart the short-hard and long-soft populations of BATSE GRBs
in the softness vs. duration plot.

\section{In the case of Swift}

GRBs observed by Swift are becoming more and more important due to
the excellent ability of Swift to detect the afterglows of the objects.
The number of bursts collected by Swift has been steadily growing up
since the launch of the satellite [10]. Would the statistical
analysis performed above be applicable to the current Swift data
set?

To answer this question, we search in literature and collect the Swift data which
contain the values of the duration and the four Swift channel
fluences. The data are available in Ref. [21] where the first Swift
BAT GRB catalog is presented. From Ref. [21] we get 222 bursts with
their $T_{90}$ as well as $f_1$ ($15-25$ keV), $f_2$
($25-50$ keV), $f_3$ ($50-100$ keV), and
$f_4$ ($100-150$ keV) being available. Compared with
Table 1 we find that this number of bursts is too small to be used
to perform the same analysis. This explains why we focus our
attention on the BATSE catalog instead of the Swift catalog.

Even though the number is so small, we manage to divide the bursts
into two groups according to their softness parameters to check
their duration distributions. The same definition of the softness
parameter is adopted, although the energy channels of Swift are not
the same as those of BATSE (also, $T_{90}$ would not be exactly the
same as that measured in BATSE, since the energy range of bursts would
play a role in determining the quantity). Those Swift bursts with
$SP \geq 0.1$ are included in the softer group, whilst the others
belong to the harder group. Duration distributions of the two groups
are shown in Figure 5. We observe that, although the statistical
significance is much less than that in the BATSE catalog due to the
small number of bursts, the same trend is observed in the figure:
the duration histogram of each group is likely to bear a
bimodality; the positions of the dip lying between the
two peaks of the bimodality of the two groups are
different, where the softer the group, the smaller value of the
duration position of the dip; when separating the bursts of each
group into short and long duration classes according to the apparent
dip, then any of the two classes occupies significantly different
percentage of the total count in different groups, e.g., the softer
the group, the larger percentage of the long duration bursts.
Comparing Figure 1 with Figure 5 one can find that the adopted statistical
analysis is hard to be applied to the current Swift data due to the
limited number of bursts.

\begin{figure}[tbp]
\begin{center}
\includegraphics[width=5in,angle=0]{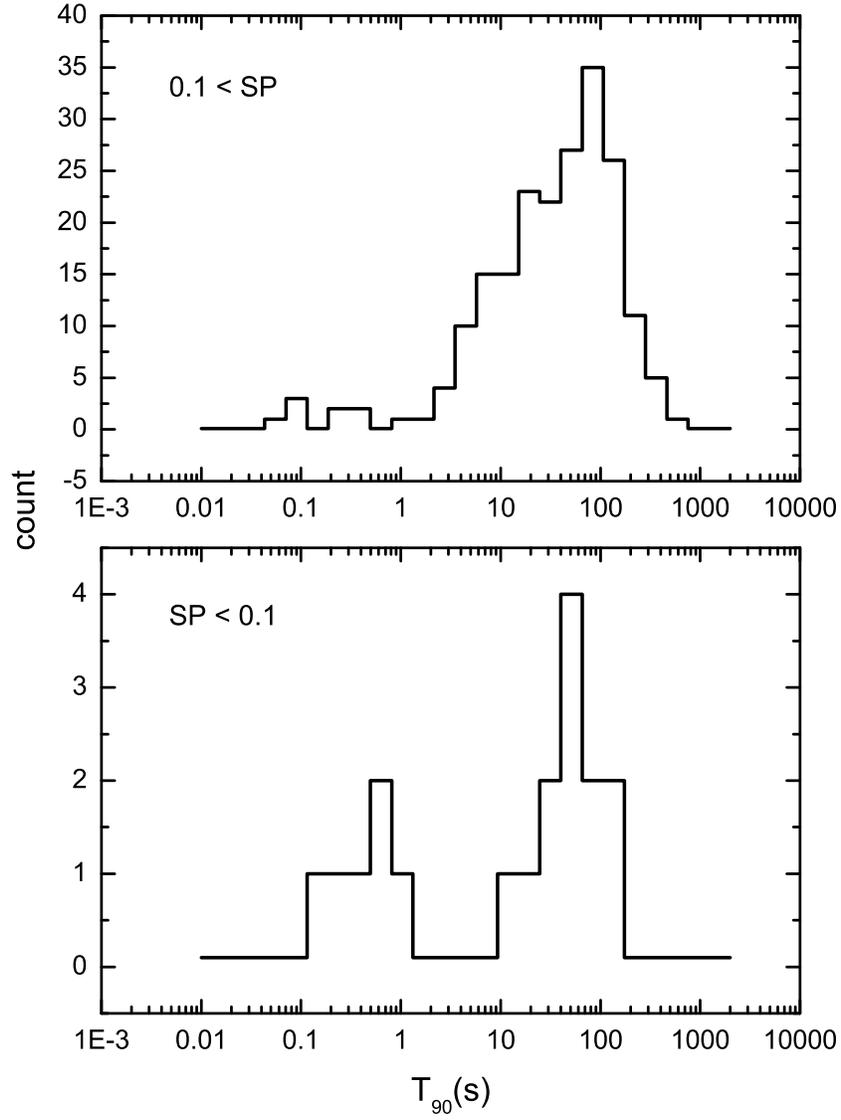}
\end{center}
\caption{Duration distributions of Swift bursts with different
softness parameters. The range of the adopted softness parameter is
mentioned at the upper left corner of each panel.} \label{Fig. 1}
\end{figure}

To check the trend in other ways, let us ignore the statistical
significance and fit a quadratic function between the two peaks in
the histogram of each panel of Figure 5 and determine the minimum of
the curve as the position of the dip. The result is displayed in
Figure 6, where as a comparison, the BATSE result shown in Figure 2 is
also presented. The same trend of the empirical function is
maintained if we rely on the two Swift data points in the figure.
However, we would like to emphasize that one should not take the two
data points so serious since the numbers adopted in Figure 5 are quite
small so that the statistical result is not so robust. In addition,
we adopt the same definition of the softness parameter to analyze
the BATSE and Swift catalogs but this definition would correspond to
at least a slightly different quantities since the energy ranges of
the corresponding channels are not the same. While this is not an
unsolvable problem currently, but the number of bursts is. We hope that
the growing Swift burst number would provide a reliable analysis in
the near future.

\begin{figure}[tbp]
\begin{center}
\includegraphics[width=5in,angle=0]{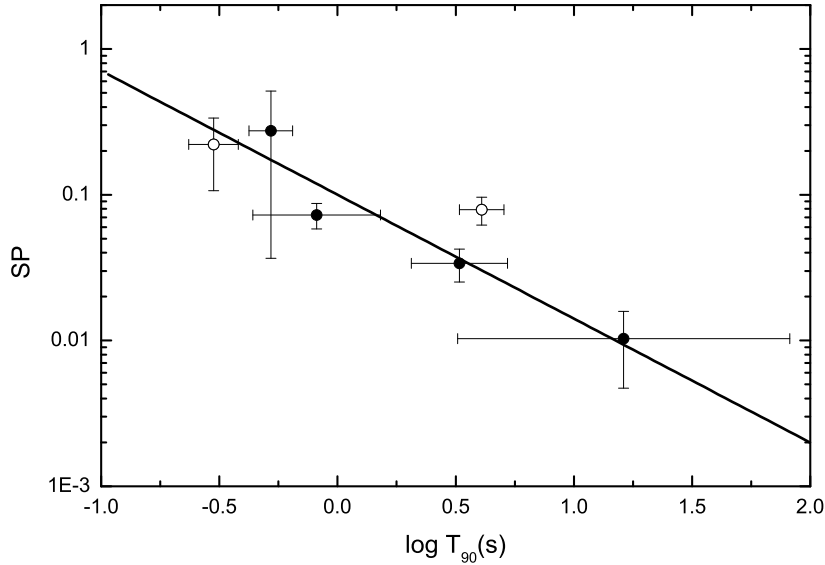}
\end{center}
\caption{Relation between the softness parameter and the position of
the duration distribution dip drawn from the Swift catalog (open
circles). Other symbols are the same as they are in Figgure 2.}
\label{Fig. 2}
\end{figure}

\section{Discussion and conclusions}

Motivated by Ref. [19]'s work, we investigate if the softness of
bursts plays a role in the conventional classification of GRBs. We
employ the BATSE catalog and define the softness parameter as the
ratio of the fluence in the first channel to the sum of the fluences
in the second, third and fourth channels. The duration distributions
of different groups of BATSE GRBs associated with distinct softness
parameters are explored. From the analysis we get an empirical curve
that can roughly set apart short and long duration bursts. The
analysis shows that the conventional classification of GRBs with the
duration of bursts is influenced by the softness of the objects.

\begin{figure}[tbp]
\begin{center}
\includegraphics[width=5in,angle=0]{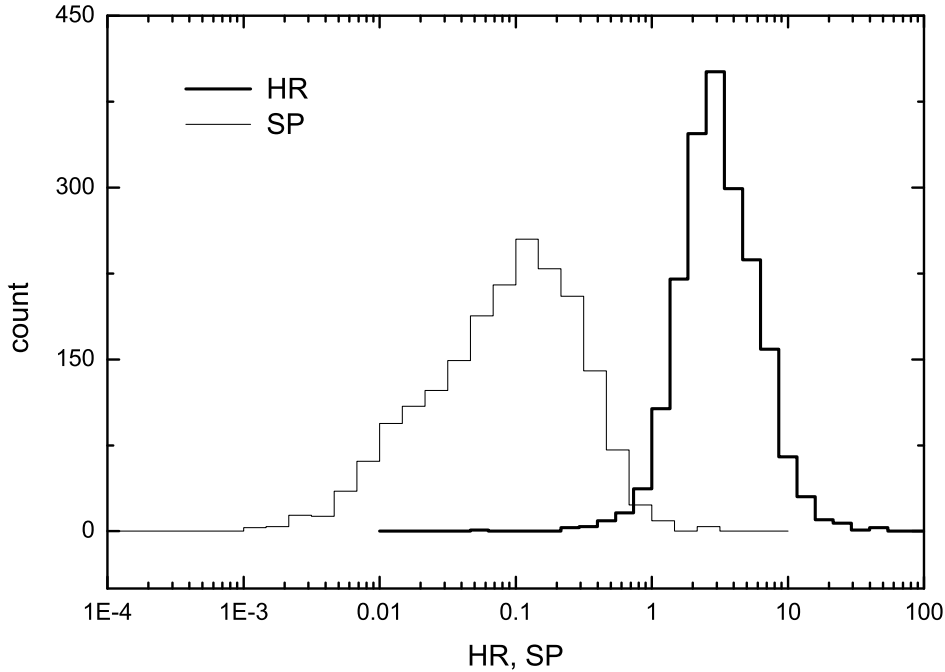}
\end{center}
\caption{Distributions of the conventional hardness ratio (the thick
line) and the softness parameter (the thin line) of the bursts in
the BATSE catalog.} \label{Fig. 1}
\end{figure}

As is known, the hardness ratio is an active factor to tell the
spectral difference between the short and long duration bursts of
GRBs. Why do we adopt the softness parameter instead of the
conventional hardness ratio to investigate the issue? As mentioned
above, it is Ref. [19]'s work on dividing XRFs, XRRs, and CGRBs with
another softness parameter ($S_X / S_\gamma$) that motivates us for
the exploration of the softness parameter as a possible factor
active in the classification of GRBs. While the role of the
conventional hardness ratio in GRB classification has been well
studied for a long time, the role of the softness parameter has not
been sufficiently explored. Another reason for doing so is
illustrated in Figure 7, where distributions of both the conventional
hardness ratio and the softness parameter are presented. We find
that, while the conventional hardness ratio spans about one
magnitude interval, the softness parameter covers approximately two
magnitude range. This suggests that, if both the conventional
hardness ratio and the softness parameter can be adopted as a
parameter to distinguish different types of burst, then the latter
must be more sensitive than the former, since the
difference of the latter is more easily noticeable.

\begin{figure}[tbp]
\begin{center}
\includegraphics[width=5in,angle=0]{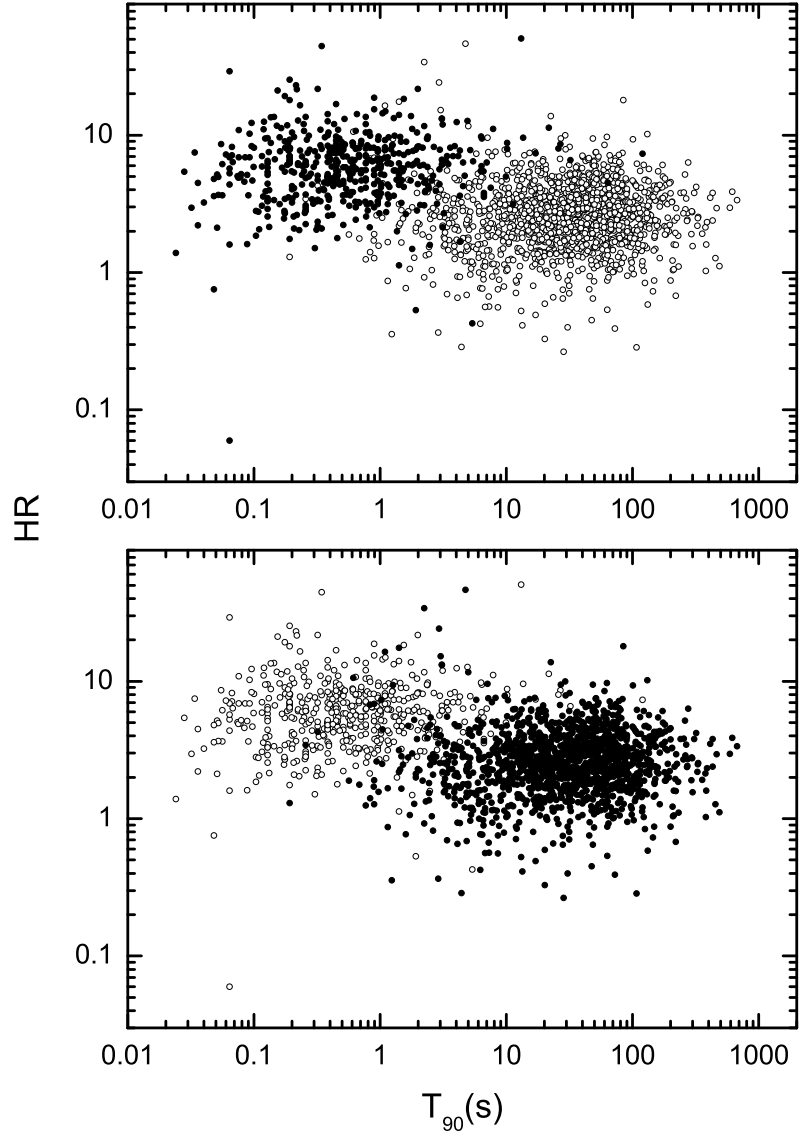}
\end{center}
\caption{Hardness ratio vs. duration plot for the short duration
bursts (filled circles in the upper panel; open circles in the lower
panel) and long duration bursts (filled circles in the lower panel;
open circles in the upper panel) divided by the function of Equation (2) from the BATSE catalog.} \label{Fig. 1}
\end{figure}

What would one get in the conventional hardness ratio vs. duration
plot, if the bursts of the BATSE catalog are divided by the
empirical curve, the function of Equation (2)? The result is shown
in Figure 8 where we find that two distinct populations are clustered
in the conventional domains of the short and long duration classes.
The well-known correlation properties between the hardness ratio and
duration of the two classical populations are maintained. That is,
for any of the two populations, the hardness ratio is not at all
correlated with the duration, while the two elements are obviously
correlated when the two populations are combined. This is not
surprising, since 84.7\% bursts of the population below the fitting
curve are the conventional short duration bursts ($T_{90}\leq 2 s$),
and 97.1\% bursts of the population above the fitting curve are the
conventional long duration bursts ($T_{90} > 2 s$). This together
with Figure 4 show that dividing BATSE bursts with the empirical curve
or with the duration criterion $T_{90} = 2 s$ would not give rise to
a statistical difference (or, the statistical properties of the two
resulting populations would not be significantly different for the
two criterions). This in turn suggests that the modification to the
conventional method (e.g., by considering different softness groups
instead of a single sample, or by adopting the empirical curve
instead of the $T_{90}=2s$ criterion) would not lead to a noticeable
physical discrepancy between the two populations.

As Figure 8 shows, the two populations divided by the empirical curve
heavily overlap in the hardness ratio vs. duration plot. This
together with the overlapping in Figures 3 and 4 imply that, the two
well-known populations are unlikely to be sharply separated merely
by quantities such as the duration, the hardness ratio and the
softness parameter, or the relations between them. To distinguish
the intrinsic short-hard and long-soft populations, other
supplemental criterions are needed. Among them, the most desirable
one is the location of bursts in the host galaxies, as observable in
some Swift bursts.

One might notice that the softness parameter adopted here is in fact
an inverted BATSE gamma-ray hardness ratio $HR_{234/1}\equiv
(f_2+f_3+f_4)/f_1$. Therefore, the fact that the dip existed in the
bimodality in the duration distribution of GRBs shifts
with the softness parameter (see Figure 1) is equivalent to the fact
that the dip position shifts with the hardness ratio $HR_{234/1}$.

What is the difference between our analysis and the conventional
one? The conventional investigation considers a single sample from a
catalog whilst our study considers different groups of softness from
the same catalog. The conventional one yields a single dip position
of the histogram of the duration distribution while our analysis
produces several dip positions which are found to vary with the
softness. The statistical methods adopted to distinguish the long
and short duration classes in both cases are exactly the same,
but the results are different. Our analysis suggests
that the conventional classification should be modified by setting
apart the two populations in different softness groups, which would
be more reasonable than doing this with a single sample. Another
possible modification is to separate the two populations in the
softness vs. duration plot with an empirical function as long as the
function is reliable enough (when the total number of bursts is much
larger than the current available one, the function so obtained
would be more robust in terms of statistics, e.g., in that case one
would get more data points rather than only 4 in Figure 2 under the
same requirements).

\begin{figure}[tbp]
\begin{center}
\includegraphics[width=5in,angle=0]{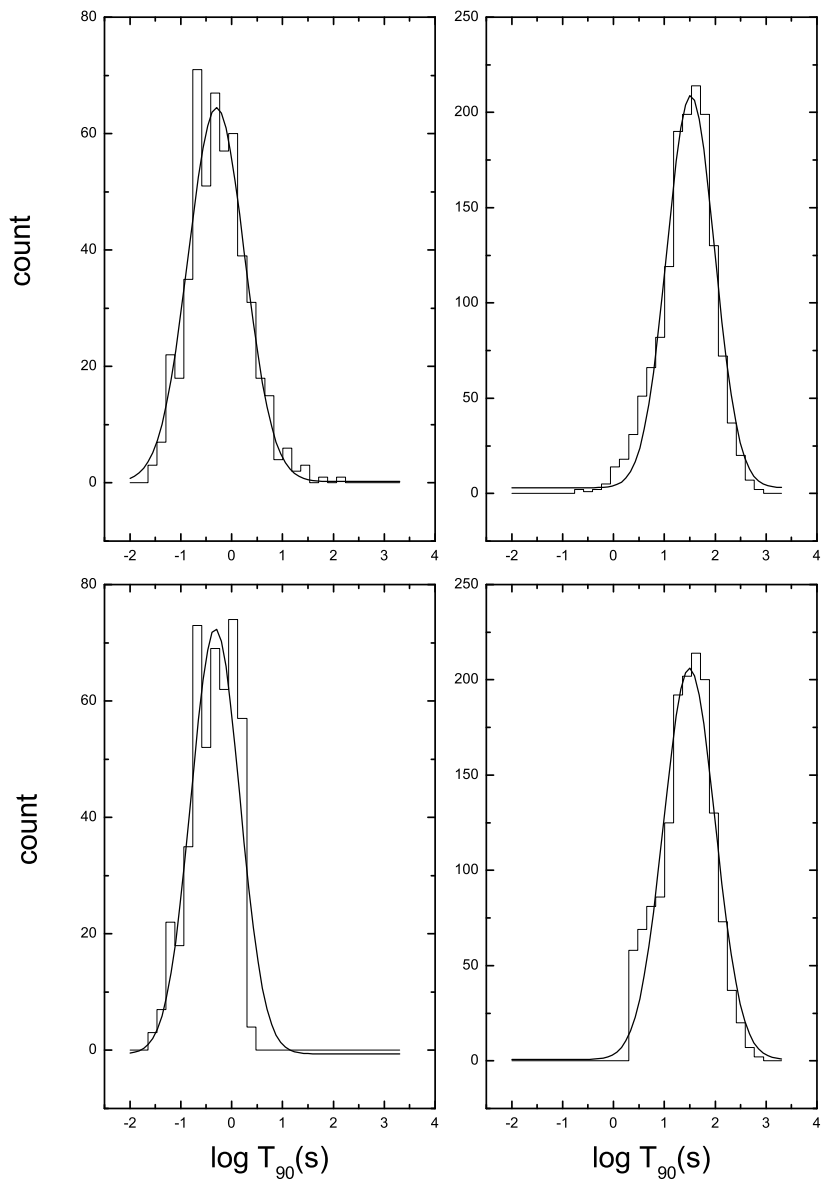}
\end{center}
\caption{Duration distributions (the thin line) of BATSE bursts
fitted by a lognormal distribution curve (the thick line) for the
conventional short (the left bottom panel; the reduced chi-square
for the fit is 85.6) and long (the right bottom panel; the reduced
chi-square for the fit is 305) duration classes, and for the newly
classified short-hard (the left top panel; the reduced chi-square
for the fit is 35.8) and long-soft (the right top panel; the reduced
chi-square for the fit is 227) populations by the empirical function
of Equation (2), respectively.} \label{Fig. 1}
\end{figure}

Assume that we take the empirical function as the criterion to
classify GRBs. If so, is the new classification statistically better
than the old one? As discussed above, as much as 84.7\% of the newly
classified short-hard bursts are the conventional short duration
bursts and as much as 97.1\% of the newly classified long-soft
bursts are the conventional long duration bursts. The new method is
only a modification to the conventional one but not a totally
different one. Therefore we cannot expect an entire difference
between bursts identified by the two kinds of classification.
However, if we believe that each of the two populations follows a
lognormal distribution of the duration of bursts, as suggested in
Ref. [17], then we can check the statistical difference of the two
classifications in this light. Presented in Figure 9 is the fitting
result of a lognormal distribution curve to the duration
distributions of BATSE bursts for the two classifications. The
statistical improvement from the new method is obviously observable.

As Figure 9 shows, compared with the short duration class, the long
duration class is more poorly fitted by a lognormal distribution.
This indicates that if the empirical function is used as a criterion
to separate the BATSE bursts, the resulting long-soft population
might contain a ``third'' subclass. As mentioned above, about one
decade ago, authors of Ref. [18] used a multivariate analysis to
discriminate between distinct classes of GRBs and found that there
exist three classes of GRBs instead of only two. The debate of the
existence of the ``third class'' has then been triggered and lasting
until the present time [22$-$28]. We suspect, if a detailed analysis based on
what is shown in Figure 9 is helpful in ending the debate, and hence
hope to see such an investigation in the near future (for example,
to explore a fit to the duration distribution of the long-soft
population with the method of the superposition of two lognormal
distributions, as that adopted in Ref. [25]).

\begin{figure}[tbp]
\begin{center}
\includegraphics[width=5in,angle=0]{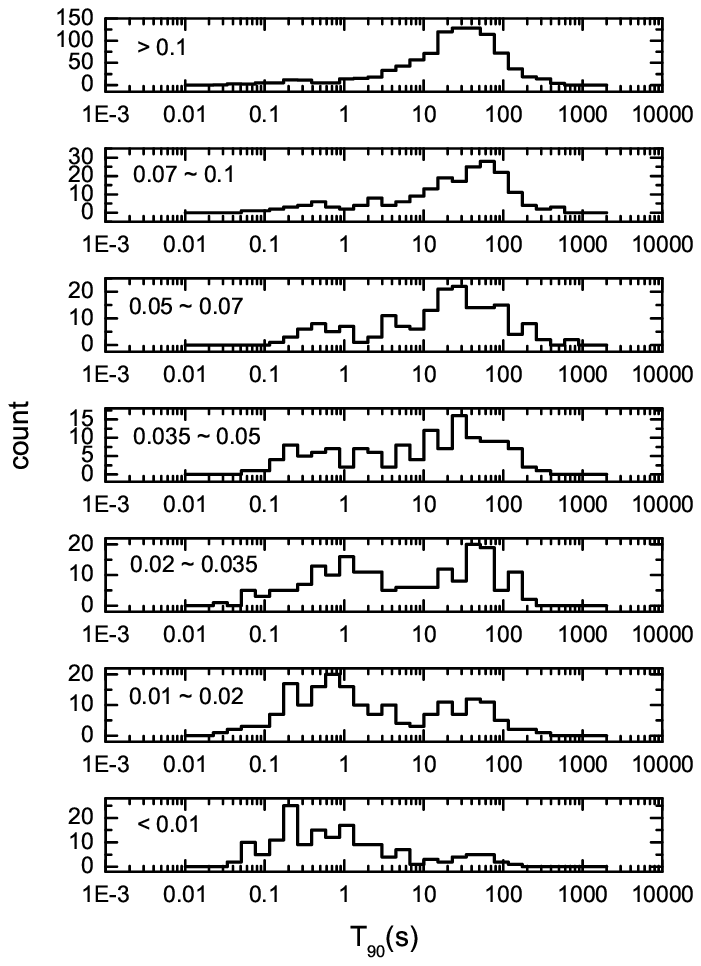}
\end{center}
\caption{Duration distributions of seven groups of the BATSE bursts
divided according to the softness parameters. Range of the adopted
softness parameter is mentioned at the upper left corner of each
panel.} \label{Fig. 1}
\end{figure}

What result would a different number of intervals of the softness
parameter produce? Obviously, larger intervals would lead to a
smaller number (smaller than 4) of data points in Figure 2, which will
not be considered here. As mentioned previously, smaller intervals
would give rise to more data points in the figure, but the
statistical significance will be harmed. For example, when we divide
the BATSE catalog into 7 instead of 4 groups, the requirement that
the maximum count of each group should be no less than 30 would no
longer be satisfied (this is the reason why we prefer to divide the
BATSE catalog into 4 groups). As a result, more obvious fluctuation
of data would be expected. The result is shown in Figure 10. Compared
with Figure 1 we find that data fluctuation is indeed more obvious in
Figure 10. Following the same method, we get an empirical curve from
Figure 10, which is $SP = (0.104 \pm 0.019) T_{90}^{-(0.95 \pm
0.11)}$. Parameters of this curve are in agreement with Equation (2)
(within the error bars, values of the fitting parameters obtained in
both cases are the same), indicating that our analysis does not rely
strongly on the choice of the number of intervals of the softness
parameter.

According to the above analysis, we reach the following
conclusions. a) When dividing BATSE bursts into several groups
according to their softness parameters, the duration distribution of
each group bears a bimodality histogram. b) Between the
two peaks of the bimodality exhibits a feature of dip
that sets apart two subgroups, shorter duration subgroup and longer
duration subgroup. c) The dip position shifts with the softness
parameter, where the softer the group, the smaller value of the
duration position of the dip. d) For groups with different softness
parameters, the short duration subgroup occupy significantly
different percentage of the total count, where the softer the group,
the smaller percentage of the total count it occupies (for the long
duration subgroup, the conclusion is the opposite). e) Deduced from the
relation between the dip position and the softness parameter we find
an empirical function that can roughly separate the short-hard and
long-soft populations.

As a primary investigation, we find a statistical difference between
the duration distributions of a single sample and multi-group
samples from the BATSE catalog. We do not know what causes this
difference. Nor are there any models that have ever predicted this.
Lacking the knowledge of the causing, the physical interpretation
associated with the proposed modification of the classification is
currently unavailable. In the same way, the phenomenon of
overlapping is currently not able to be explained. Other independent
investigations are needed to understand the new findings as well as
the overlapping phenomenon.

\acknowledgments

This work is supported in part by the National Natural Scientific
Foundation of China (10633010, 10573005, 10747001) and the 973
project (No. 2007CB815405). We also thank the financial support from
the Guangzhou Education Bureau and Guangzhou Science and Technology
Bureau.\\

1  Kouveliotou C, Meegan C A, Fishman G J, et al. Identification of
two classes of gamma-ray bursts. Astrophys J, 1993, 413(2):
L101-L104

2  Fishman G J, Meegan C A. Gamma-Ray Bursts. Annu Rev Astron
Astrophys, 1995, 33: 415-458

3  Paciesas W S, Meegan C A, Pendleton G N, et al. The Fourth BATSE
Gamma-Ray Burst Catalog (Revised). Astrophys J Suppl Ser, 1999,
122(2): 465-495

4  Qin Y-P, Xie G-Z, Xue S-J, et al. The Hardness-Duration
Correlation in the Two Classes of Gamma-Ray Bursts. Publ Astron Soc
Japan, 2000, 52: 759-761

5  Woosley S E. Gamma-ray bursts from stellar mass accretion disks
around black holes. Astrophys J, 1993, 405(1): 273-277

6  Paczynski B. Are Gamma-Ray Bursts in Star-Forming Regions?
Astrophys J, 1998, 494: L45-L48

7  MacFadyen A I, Woosley S E. Collapsars: Gamma-Ray Bursts and
Explosions in ``Failed Supernovae''. Astrophys J, 1999, 524(1):
262-289

8  Paczynski B. Gamma-ray bursters at cosmological distances.
Astrophys J, 1986, 308: L43-L46

9  Eichler D, Livio M, Piran T, et al. Nucleosynthesis, neutrino
bursts and gamma-rays from coalescing neutron stars. Nature, 1989,
340: 126-128

10  Gehrels N, Chincarini G, Giommi P, et al. The Swift Gamma-Ray
Burst Mission. Astrophys J, 2004, 611(2): 1005-1020

11  Fruchter A S, Levan A J, Strolger L, et al. Long $\gamma$-ray
bursts and core-collapse supernovae have different environments.
Nature, 2006, 441(7092): 463-468

12  Stanek K Z, Matheson T, Garnavich P M, et al. Spectroscopic
Discovery of the Supernova 2003dh Associated with GRB 030329.
Astrophys J, 2003, 591(1): L17-L20

13  Hjorth J, Sollerman J, Moller P, et al. A very energetic
supernova associated with the $\gamma$-ray burst of 29 March 2003.
Nature, 2003, 423(6942): 847-850

14  Barthelmy S D, Chincarini G, Burrows D N, et al. An origin for
short $\gamma$-ray bursts unassociated with current star formation.
Nature, 2005, 438(7070): 994-996

15  Berger E, Price P A, Cenko S B, et al. The afterglow and
elliptical host galaxy of the short $\gamma$-ray burst GRB 050724.
Nature, 2005, 438(7070): 988-990

16  Hjorth J, Watson D, Fynbo J P U, et al. The optical afterglow of
the short $\gamma$-ray burst GRB 050709. Nature, 2005, 437(7060):
859-861

17  McBreen B, Hurley K J, Long R, et al. Lognormal Distributions in
Gamma-Ray Bursts and Cosmic Lightning. Mon Not R Astron Soc, 1994,
271: 662-666

18  Mukherjee S, Feigelson E D, Jogesh Babu G, et al. Three Types of
Gamma-Ray Bursts. Astrophys J, 1998, 508(1): 314-327

19  Sakamoto T, Lamb D Q, Kawai N, et al. Global Characteristics of
X-Ray Flashes and X-Ray-Rich Gamma-Ray Bursts Observed by HETE-2.
Astrophys J, 2005, 629(1): 311-327

20  Sakamoto T, Hullinger D, Sato G, et al. Global Properties of
X-Ray Flashes and X-Ray-Rich Gamma-Ray Bursts Observed by Swift.
Astrophys J, 2008, 679(1): 570-586

21  Sakamoto T, Barthelmy S D, Barbier L, et al. The First Swift BAT
Gamma-Ray Burst Catalog. Astrophys J Suppl Ser, 2008, 175(1):
179-190

22  Horv¨¢th I A. Third Class of Gamma-Ray Bursts? Astrophys J,
1998, 508(2): 757-759

23  Hakkila J, Haglin D J, Pendleton G N, et al. Gamma-Ray Burst
Class Properties. Astrophys J, 2000, 538(1): 165-180

24  Balastegui A, Ruiz-Lapuente P, Canal R. Reclassification of
gamma-ray bursts. Mon Not R Astron Soc, 2001, 328(1): 283-290

25  Horv¨¢th I. A further study of the BATSE Gamma-Ray Burst
duration distribution. Astron Astrophys, 2002, 392: 791-793

26  Rajaniemi H J, Mahonen P. Classifying Gamma-Ray Bursts using
Self-organizing Maps. Astrophys J, 2002, 566(1): 202-209

27  Chattopadhyay T, Misra R, Chattopadhyay A K, et al. Statistical
Evidence for Three Classes of Gamma-Ray Bursts. Astrophys J, 2007,
667(2): 1017-1023

28  Henry J P. A Measurement of the Density Parameter Derived from
the Evolution of Cluster X-Ray Temperatures. Astrophys J, 1997, 489:
L1-L5

\end{document}